\documentclass{jfm}

\usepackage{graphicx}
\usepackage{natbib}
\usepackage{math_pack}
\usepackage{color}
\usepackage{subfigure}
\usepackage{amsmath}
\usepackage{amssymb}
\usepackage{math_pack}

\ifCUPmtlplainloaded \else
  \checkfont{eurm10}
  \iffontfound
    \IfFileExists{upmath.sty}
      {\typeout{^^JFound AMS Euler Roman fonts on the system,
                   using the 'upmath' package.^^J}%
       \usepackage{upmath}}
      {\typeout{^^JFound AMS Euler Roman fonts on the system, but you
                   dont seem to have the}%
       \typeout{'upmath' package installed. JFM.cls can take advantage
                 of these fonts,^^Jif you use 'upmath' package.^^J}%
      }
  \else
  \fi
\fi

\ifCUPmtlplainloaded \else
  \checkfont{msam10}
  \iffontfound
    \IfFileExists{amssymb.sty}
      {\typeout{^^JFound AMS Symbol fonts on the system, using the
                'amssymb' package.^^J}%
       \usepackage{amssymb}%
         
       \let\ge=\geqslant  
      }{}
  \fi
\fi

\ifCUPmtlplainloaded \else
  \IfFileExists{amsbsy.sty}
    {\typeout{^^JFound the 'amsbsy' package on the system, using it.^^J}%
     \usepackage{amsbsy}}
    {\providecommand\boldsymbol[1]{\mbox{\boldmath $##1$}}}
\fi

\newcommand\taub{\boldsymbol{\tau}}

\newsavebox{\astrutbox}
\sbox{\astrutbox}{\rule[-5pt]{0pt}{20pt}}

\definecolor{cinnamon}{rgb}{0.82, 0.41, 0.12}

\title[Reduced particle settling speed in turbulence]{Reduced particle settling speed in turbulence}

\author[W. Fornari, F. Picano, G. Sardina and L. Brandt]%
{Walter Fornari$^1$%
  \thanks{Email address for correspondence: fornari@mech.kth.se},
Francesco Picano$^2$,  Gaetano Sardina$^1$ \\
and Luca Brandt$^1$}

\affiliation{$^1$Linn\'e Flow Centre and Swedish e-Science Research Centre (SeRC), \\ KTH Mechanics,
SE-100 44 Stockholm, Sweden\\[\affilskip]
$^2$Department of Industrial Engineering, University of Padova, \\ Via Venezia 1, 35131 Padua, Italy}

\pubyear{2010}
\volume{650}
\pagerange{119--126}
\date{?; revised ?; accepted ?. - To be entered by editorial office}
\begin{document}

\maketitle

\begin{abstract}

We study the settling of finite-size rigid spheres in sustained homogeneous isotropic turbulence (HIT) by direct numerical 
simulations using an immersed boundary method to account for the dispersed solid phase. We study semi-dilute 
suspensions at different Galileo numbers, $Ga$. {The Galileo number is the ratio between buoyancy and viscous forces, and is} 
here varied via the solid-to-fluid density ratio $\rho_p/\rho_f$. {The focus is on particles that are slightly heavier than the fluid.} 
We find that in HIT, the mean settling speed is less than that
in quiescent fluid; in particular it reduces by $6\%$-$60\%$  with respect to the terminal velocity of an isolated sphere in quiescent fluid
as the ratio between the latter  and the turbulent velocity fluctuations $u'$ is decreased. Analysing the fluid-particle relative 
motion, we find that the mean settling speed is progressively reduced while reducing $\rho_p/\rho_f$ due to the increase of the vertical 
drag induced by the particle cross-flow velocity. Unsteady effects contribute to the mean overall drag by about $6\%$-$10\%$. 
The probability density functions of particle velocities and accelerations
reveal that these are closely related to the features of the turbulent flow. The particle mean-square displacement in the 
settling direction is found to be similar for all $Ga$ if time is scaled by $(2a)/u'$ (where $2a$ is the particle diameter and $u'$ is 
the turbulence velocity root mean square).
\end{abstract}


\section{Introduction}

The gravity-driven motion of solid particles in a viscous fluid is a relevant process 
in a wide number of environmental and engineering applications. Among these we recall 
volcanic eruptions, fluidized beds, soot particle dispersion, rain droplets, snow and settling of micro-organisms such as plankton.\\
The settling process may occur in quiescent fluids or in already-turbulent flows. In the latter case 
the settling dynamics, which depends on the solid-to-fluid density ratio $\rho_p/\rho_f$, 
the solid volume fraction $\phi$ and on the Galileo number $Ga$ (i.e. the ratio between buoyancy and viscous forces), is further 
complicated by the interaction among the particles and the turbulent eddies. 

The vast majority of previous investigations focused on the settling of particles smaller or at least comparable in size to the 
Kolmogorov lengthscale, $\eta$. In these conditions, turbulence can either enhance, reduce or inhibit the settling. 
As shown by \citet{squires1991}, small inertial particles tend to be expelled from the vortex core 
and accumulate in regions of low vorticity 
and high strain rate. Owing to this and to gravitational settling, particles are often swept into regions of downdrafts (the so-called preferential 
sweeping or fast-tracking). Thus, the particle mean settling velocity increases, as first observed in simulations in random 
flows \citep{maxey1987} and in turbulence \citep{wang1993}, and later confirmed by experiments \citep{nielsen1993,aliseda2002,yang2003,yang2005}.

A reduction in mean settling velocity, on the other hand, has also been observed both in experiments \citep{murray1970,
nielsen1993,yang2003,kawanisi2008} and numerical simulations \citep{wang1993,good2014}. Note, however, that reduction of the  mean settling velocity can only be observed in direct numerical 
simulations (DNS) of sub-Kolmogorov particles if nonlinear drag corrections are employed \citep[i.e.\ for a finite particle Reynolds number, $Re_p$, as shown by][]{good2014}. 
Reduced settling speeds are observed when particles oversample upward flow and not downward motions as in the case of preferential sweeping. \citet{nielsen1993} suggested  that fast-falling particles 
need longer times to cross regions of upward flow (a 
phenomenology usually referred to as loitering), the more so if the particle settling speed is of the order of the turbulent velocity 
fluctuations, $u'$. 
\citet{good2014} performed a series of experiments and numerical simulations 
and found that reduction of the mean settling velocity occurs when the ratio $\tau_p g/u'$ (where $\tau_p=2(\rho_p/\rho_f-1)a^2/(9 \nu)$ is the particle 
relaxation time with $a$ the particle radius and $\nu$ the fluid viscosity) is greater than one (i.e.\ when the particle terminal 
velocity is larger than the turbulent velocity fluctuations). {Heuristically it can be said that when} $\tau_p g$ 
(the Stokes settling velocity) is sufficiently high, the particles fall along almost straight vertical paths, their horizontal velocity fluctuations are 
weak and hence they are unable to side step the turbulent eddies: fast-tracking is suppressed and the mean settling velocity reduces due to a drag increase  related to finite Reynolds number.\\
\indent Thus far, just few studies consider the settling of finite-size particles in turbulent environments. 
The experiments by \citet{byron2015} investigate the settling of Taylor-scale particles 
using refractive-index-matched hydrogel particles and particle image velocimetry (PIV) and show that particles with quiescent settling 
velocities of the same order of the turbulence root-mean-square (r.m.s.) velocity fall on average $40\%-60\%$ more slowly in turbulence 
(depending on their density and shape).
Previous numerical studies had focused mostly on settling in quiescent environments \cite[see, for example][]{yin2007,uhlmann2014,zaidi2014}, or 
on the dynamics of neutrally buoyant particles in homogeneous isotropic turbulence (HIT) \citep{homann2010}. Recently, \citet{fornari2015} 
compared the settling of spheres 
in quiescent and sustained HIT. In this study, the sphere radius was chosen to be about six Kolmogorov lengthscales,  
the density ratio $\rho_p/\rho_f=1.02$ and the ratio between the quiescent settling velocity and the turbulence r.m.s. velocity $u'$ was chosen to be about
$3.3$. 
In dilute conditions, the particle mean settling velocity reduces by 
about $4\%$ in still fluid and  by about $12\%$ in turbulence when compared with the terminal velocity $V_t$ of an isolated particle. This reduction is attributed to unsteady phenomena such as vortex shedding (absent in the quiescent cases), 
and to the modification of the particle wakes by the turbulence \citep[see also][]{bagchi2003,homann2013}.\\
In the present study, we investigate the effect of the Galileo number, $Ga$, 
on the settling in a turbulent environment. 
The background sustained homogeneous isotropic turbulent flow has a nominal 
Reynolds number based on the Taylor microscale $Re_{\lambda}$ of about $90$. By varying the Galileo number, via the density ratio $\rho_p/\rho_f$, we control the ratio between the terminal velocity $V_t$ and the {turbulent} 
velocity fluctuations $u'$. We show that the 
reduction in mean settling velocity 
increases from about $10\%$ to $55\%$ as the Galileo number $Ga$ (i.e. $V_t/u'$) is reduced. 
Analysing the mean forces acting 
on the particles, we attribute the significant reduction in mean settling velocity observed at the lower $Ga$ to the increase of the vertical component of the drag originating from the horizontal components of the particle relative velocity. 

\section{Set-up and Methodology}\label{sec:method}

Sedimentation of a dilute suspension is considered in a computational domain with periodic
boundary conditions in the $x$, $y$ and $z$ directions, with gravity acting in the positive $z$ direction.
The computational box has size $32a\times32a\times320a$ and  the volume fraction $\phi=0.5\%$, corresponding to
$391$ particles; these are initially  randomly distributed in the computational volume with zero velocity and rotation.
We consider non-Brownian rigid spherical particles, slightly heavier than the suspending fluid with density ratios $\rho_p/\rho_f=1.00035$, $1.0034$, $1.020$ 
and $1.038$. The parameter governing the settling is the Galileo number $Ga=\sqrt{\left({\rho_p}/{\rho_f}-1\right) g (2a)^3}/{\nu}$,
the non-dimensional number that quantifies the importance of the gravitational forces acting on the particle with respect to viscous forces. 
For the different density ratios $\rho_p/\rho_f$ considered here we have $Ga=19$, $60$, $145$ and $200$. 

To generate and sustain an isotropic and homogeneous turbulent flow field, a random forcing is applied to the first wavenumber in the directions perpendicular to 
gravity, and to the tenth wavenumber in the settling direction. Since in the settling direction the box length is $10$ times that in the other directions, 
forcing the tenth wavenumber is equivalent to forcing the first wavenumber in a cube of size $32a\times32a\times32a$. The forcing is $\delta$-correlated 
in time and of fixed amplitude \citep{vincent1991}. This forcing  induces a turbulent flow with Reynolds number based on the 
Taylor microscale, $Re_{\lambda}=\lambda u'/\nu=90$ (where $u'$ is the fluctuating velocity r.m.s., 
 $\lambda=\sqrt{15\nu u'^2/\epsilon}$ the transverse Taylor length scale and $\epsilon$ the dissipation). 
The ratio between the Kolmogorov lengthscale $\eta=(\nu^3/\epsilon)^{1/4}$ and the grid spacing ($\eta/\Delta x$)
is approximately $1.3$ while the particle diameter is approximately $12\eta$. The parameters of the turbulent flow field are summarized in table~\ref{tab:turb}.

\begin{table}
  \begin{center}
\def~{\hphantom{0}}
  \begin{tabular}{ccccccccc}
      $(2a)/\eta$  &   $\lambda/(2a)$ & $Re_{\lambda}$ & $\lambda/\eta$ & $L_{\epsilon}/\eta$ & $u'/u_{\eta}$ & $L_x/L_{\epsilon}$ & $L_z/L_{\epsilon}$\\[3pt]
       11.9   &  1.56 & 90 & 18.6 & 120 & 4.76 & 1.6 & 16 \\
  \end{tabular}
  \caption{Turbulent flow parameters pertaining the present DNS, where $L_{\epsilon}$ is the integral lengthscale and $u_{\eta}$ is the Kolmogorov velocity scale. 
The box size in the directions perpendicular and parallel to gravity is denoted by $L_x$ and $L_z$.}
  \label{tab:turb}
  \end{center}
\end{table}

Each Galileo number $Ga$ also defines a different value of the ratio between the terminal velocity $V_t$ (i.e. the settling velocity of a single particle in 
quiescent fluid) and the turbulent velocity fluctuations. This parameter directly influences the average settling as noted by \citet{nielsen1993} and \citet{byron2015} among others. 
For the four cases considered here this ratio attains the values $V_t/u'=0.19$, $0.99$, $3.38$ and $4.81$.

The simulations have been performed using the immersed boundary method originally developed by \citet{breugem2012}; this fully models the coupling between the 
solid and fluid phases. The 
flow is evolved according to the incompressible Navier-Stokes equations, whereas the particle motion is governed  by the Newton-Euler Lagrangian equations for the particle 
centroid linear and angular velocities. Using the immersed boundary method, the boundary condition at the moving fluid/solid interfaces is modelled by an additional force 
on the right-hand side of the Navier-Stokes equations, making it possible to discretize the computational domain with a fixed staggered mesh on which 
the fluid phase is evolved using a second-order finite-difference scheme. Time integration is performed by a third-order Runge-Kutta scheme combined with pressure 
correction at each sub-step. When the distance between two particles becomes smaller than twice the mesh size, lubrication models based on Brenner's asymptotic 
solution \citep{brenner1961} are used to correctly reproduce the interaction between the particles. A soft-sphere collision model is used to account for collisions 
between particles. An almost elastic rebound is ensured with a restitution coefficient set at $0.97$. 
A cubic mesh with eight points per particle radius is used for the results presented, which corresponds to $256\times256\times2560$ grid points. 
This resolution is a good compromise in terms of accuracy and computational cost as shown in previous publications \citep{breugem2012,lambert2013,picano2015,lash2016,fornari2016,fornari2015}, 
wherein more details and validations of the numerical code are provided. 
Note finally that zero total volume flux is imposed in the simulations. 

When studying settling in a turbulent flow, the fluid phase is evolved for approximately six eddy turnover times before adding the solid phase. Statistics are 
collected after an initial transient phase so that the difference between the statistics presented here and those computed from half the samples is below $1\%$ for 
the first and second moments. The transient is approximately nine integral time scales ($T_{\epsilon}=L_{\epsilon}/u'$) in HIT and at least $100 \tau_p$ in the 
quiescent cases. After these transient periods, 
velocities and accelerations oscillate on average with a constant amplitude around the mean. In the following 
we will use $\vec U$ and $\vec V$ for the fluid and particle velocities.

\section{Results}
\subsection{Particle statistics}

The most striking result of our study is that in a turbulent flow, as the Galileo number is reduced, the mean settling speed $\langle V_z \rangle_{p,t}$ 
becomes significantly smaller than the terminal velocity of a single particle in still fluid, $V_t$. The notation $\langle \cdot \rangle_{p,t}$ denotes 
averaging over the total number of particles and time.

\begin{figure}
  \centering
  \subfigure{%
    \includegraphics[scale=0.34]{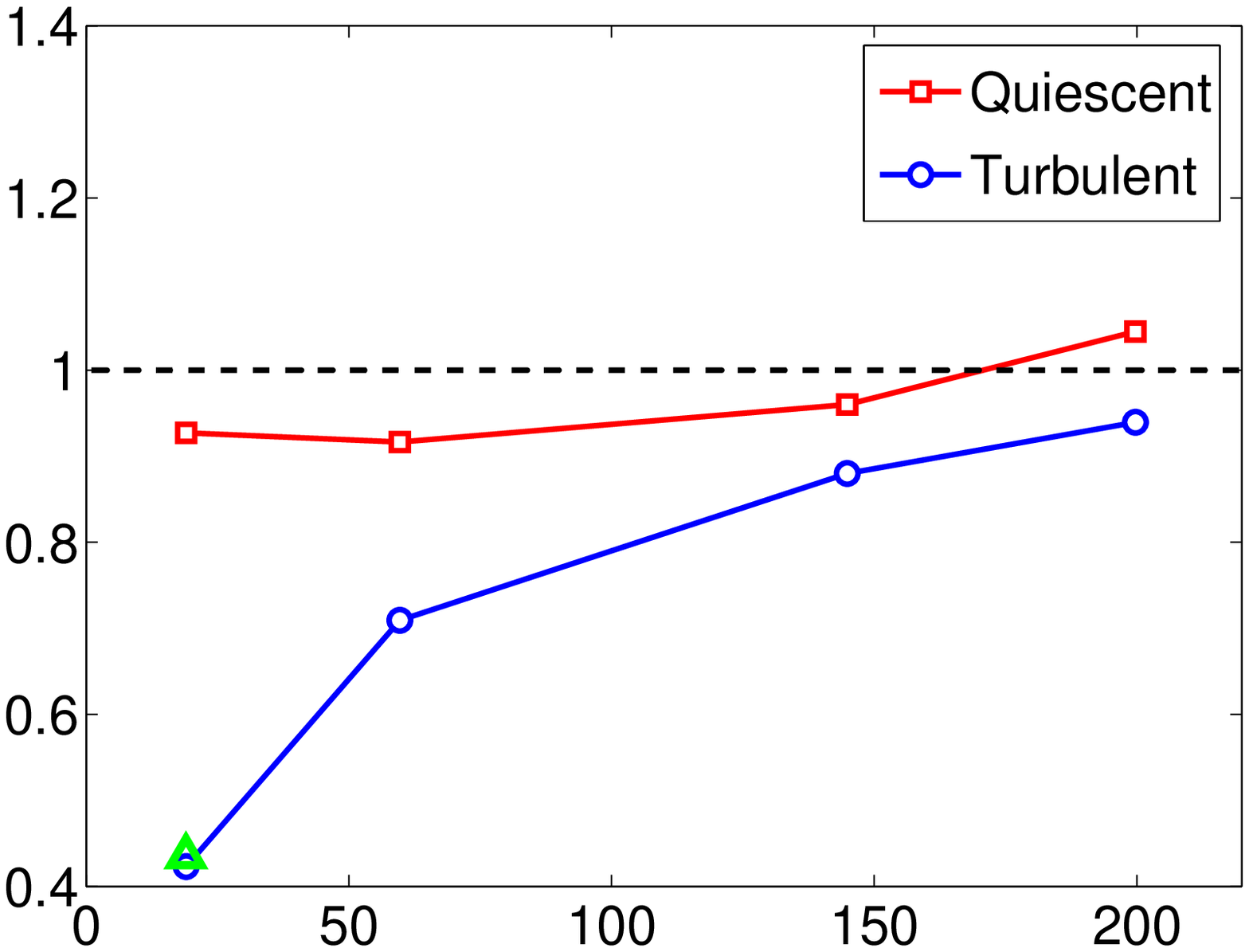}   
     \put(-194,120){{\large a)}}
     \put(-194,50){\rotatebox{90}{\large $\langle V_{z}\rangle_{p,t}/V_t$}}
     \put(-104,-5){{\large $Ga$}} 
     }%
  \subfigure{%
    \includegraphics[trim=0.5cm 6.5cm 0.0cm 7.0cm, clip=true, scale=0.33]{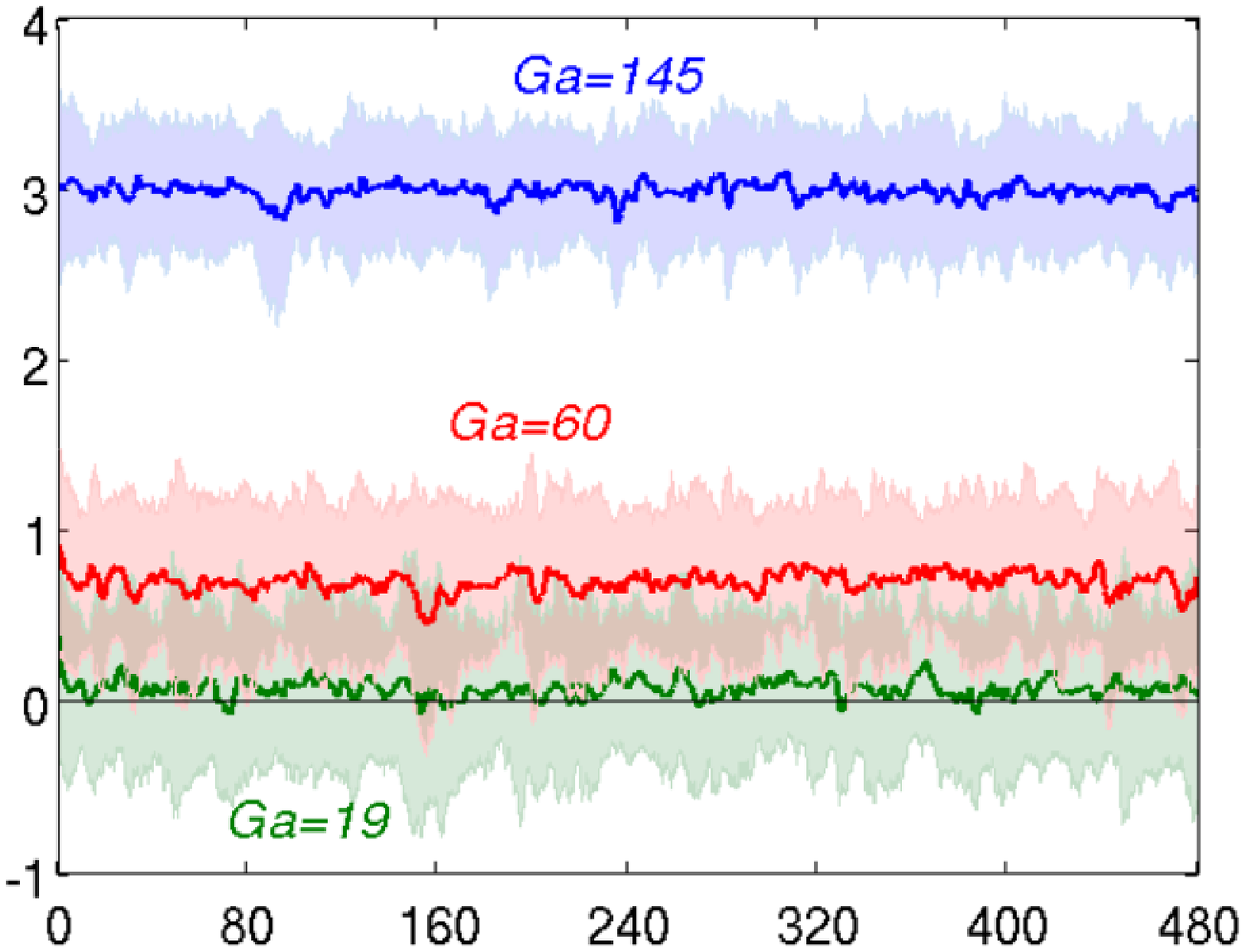}
       \put(-194,120){{\large b)}}
     \put(-194,50){\rotatebox{90}{\large $\langle V_{z}\rangle_p/u'$}}
     \put(-115,-5){{\large $t\, u'/(2a)$}}
}%
\caption{(a) Mean settling velocity $\langle V_{z}\rangle_{p,t}$ as a function of $Ga$ for both quiescent and turbulent cases. The green 
triangle shows the result obtained via $2$ additional simulations with $Ga=19$ and $\rho_p/\rho_f=1.02$ (same $\rho_p/\rho_f$ as for $Ga=145$ 
but with smaller $g$). (b) 
Evolution in time of $\langle V_{z}\rangle_p/u'$ for $Ga=19$, $60$ and $145$. 
The shaded zones represent at each time the range of variation of $\langle V_{z}\rangle_p/u'$.}
\label{fig:velsed}
\end{figure}

The mean settling speed $\langle V_z \rangle_{p,t}$ normalized by $V_t$, is shown as a function of $Ga$ in figure~\ref{fig:velsed}(a) for both quiescent and 
turbulent cases.
In quiescent fluid, $\langle V_z \rangle_{p,t}$ is smaller than the terminal velocity $V_t$ of an isolated 
particle due to the hindering effect \citep{yin2007,guazzelli2011} for the three smallest Galileo numbers.  
The ratio $\langle V_z \rangle_{p,t}/V_t \approx 0.92$ for $Ga=19$ and $60$, increases to $0.96$ for $Ga=145$ and becomes larger than unity, $1.05$, for $Ga=200$.  
This is associated with the formation of particle clusters that settle faster than isolated 
particles, as documented in \citet{uhlmann2014}. Indeed, the probability density function ($p.d.f.$) of $\langle V_z \rangle_{p,t}/V_t$ for $Ga=200$ is 
highly skewed toward settling speeds greater than the modal value (not shown). 

The mean settling speed $\langle V_z \rangle_{p,t}$ is always lower in sustained HIT than in quiescent fluid and for the Galileo 
numbers here investigated here, smaller than the terminal velocity $V_t$. 
The reduction of $\langle V_z \rangle_{p,t}$ with respect to the quiescent cases is $55\%$, $23\%$, $9\%$ and $10\%$ for $Ga=19,$ $60$, $145$ and $200$. 
As $V_t/u'$ is reduced, the reduction in average settling speed 
drastically increases. Interestingly, the settling speed of heavy sub-Kolmogorov particles reduces instead when increasing the ratio $V_t/u'$. The two different 
mechanisms will be discussed below. 

It is worth noting that we could have changed the Galileo number $Ga$ via the gravity $g$ while keeping $\rho_p/\rho_f$ constant. To check the effect of 
varying $Ga$ via $g$, we have performed additional simulations at $Ga=19$ and $\rho_p/\rho_f=1.02$ (the density ratio used for $Ga=145$). 
We consider  a single sphere settling in quiescent fluid as well as a suspension settling in a turbulent flow ($\phi=0.5\%$) and find 
that the mean settling speed $\langle V_z \rangle_{p,t}$ is about $43\% V_t$ while  
$\langle V_z \rangle_{p,t} \sim 42\% V_t$ when the same $Ga$ is obtained with $\rho_p/\rho_f=1.00035$. 
Hence, the results discussed here can be extended to suspensions of slightly buoyant 
settling spheres with different $Ga$ but constant $\rho_p/\rho_f$. 
The normalized mean settling speed $\langle V_z \rangle_{p,t}/V_t$ obtained from this 
last set  of simulations is also reported in figure~\ref{fig:velsed}(a) (green triangle).

The time evolution of $\langle V_z \rangle_p/u'$, the velocity component in the direction of gravity averaged over the total number of spheres, 
is reported after the initial transients in figure~\ref{fig:velsed}(b) for the three lowest $Ga$ considered. 
The shaded zones around each 
curve represent the instantaneous range of variation of $\langle V_z \rangle_p/u'$, which is larger in the cases at lower $Ga$. At $Ga=19$, 
when the ratio $\langle V_z \rangle_p/u'$ is closest to zero ($\sim 0.07$), the turbulence intensity is sufficiently high for particles to move in the direction opposite to gravity. Since the turbulent velocity fluctuations are considerably larger 
than the mean settling speed, particles are also subject to strong lateral motions.
To understand the significant reduction in $\langle V_z \rangle_{p,t}$ we consider the balance of the mean forces acting on the particles.

\subsection{Force analysis}

The equation of motion for a spherical particle settling due to gravity reads
\begin{equation}
\label{eq_sph}
\frac{4}{3} \pi a^3 \rho_p \td{\vec V}{t} = \frac{4}{3} \pi a^3 (\rho_p - \rho_f) \vec g + \oint_{\partial \mathcal{V}_p}^{} \vec \taub \cdot \vec n\, dS
\end{equation}
where the integral is over the surface of the sphere $\partial \mathcal{V}_p$, $\vec n$ is the outward normal and $\vec \taub$ is the fluid stress. 
As usually done in aerodynamics, equation~(\ref{eq_sph}) can be rewritten as
\begin{equation}
\label{eq_sph2}
\frac{4}{3} \pi a^3 \rho_p \td{\vec V}{t} = \frac{4}{3} \pi a^3 (\rho_p - \rho_f) \vec g - \vec D
\end{equation}
where $\vec D$ is the drag acting on the particle \citep[see also][]{fornari2015}. This drag term can be further expressed as the sum of two contributions: 
the first depends only on the particle Reynolds number, $Re_p$, while the second accounts for various non-stationary effects (such as history effects and hindering).
For sub-Kolmogorov particles with Reynolds numbers $Re_p < 1$ in unsteady non-uniform flows, the correct form of $\vec D$ was derived by \citet{maxey1983} as the sum of  Stokes drag, pressure gradient, added mass and Basset history forces.
 
Ensemble averaging {eq.~(\ref{eq_sph2})} over time and the number of particles we can isolate single contributions to the overall drag. The steady-state average equation 
projected along the direction of gravity reads
\begin{equation}
\label{eq_sph_av}
0 = \frac{4}{3} \pi a^3 (\rho_p - \rho_f) g -  F_D^S - F_D^U.\\
\end{equation}
where $F_D^S$ and $F_D^U$ are the mean contributions to the overall drag due to steady nonlinear effects and to unsteady effects (such 
as those due to the history force and hindering).\\
The particle Reynolds number is defined as $Re_p=2a|\vec U_{r}|/\nu$, where $\vec U_{r}$ is the relative velocity between particles and fluid. The term 
$F_D^S$ depends only on $Re_p$ and can be written as 
\begin{equation}
\label{eq_sph_fd}
F_D^S = \frac{1}{2} \rho_f \pi a^2 \langle |\vec U_{r}| U_{r,z} C_{D_0}(Re_p) \rangle
\end{equation}
where $\pi a^2$ is the particle reference area and $C_{D_0}(Re_p)$ the steady drag coefficient. 
The first term on the right-hand side of equation~(\ref{eq_sph_av}) is known, while $F_D^S$ can be 
calculated from the relative velocity $\vec U_r$ and the steady drag coefficient $C_{D_0}$.

To evaluate $\vec U_r$ from the present simulations, we consider spherical shells surrounding each particle, inspired by the 
works of \citet{bellani2012,cisse2013,kidan2013}. The relative velocity
is calculated as the difference between the particle velocity and the fluid velocity averaged over the volume of the shell
of inner radius $\Delta$. 
The thickness of the shell is $\delta=(2a)/8$ while $\Delta$ is $3.5a$ for all $Ga$. Here $\Delta$ is chosen large enough so
that the shell is well outside the boundary layer at particle surface (due to no-slip and no-penetration conditions) and small enough 
for the fluid motion to be still correlated to that of the particle \citep[see the discussion in][]{fornari2015}. 
Once the relative velocity $\vec U_r$ is known, the steady drag coefficient $C_{D_0}$ can be found by means of empirical 
formulae. 
Among the different expressions 
for the nonlinear 
correction to the Stokes drag that can be found in the literature \citep{schil1935,di1999,yin2007}, we follow \cite{yin2007}:
\begin{equation}
\label{cdd}
C_{D_0} = ({24}/{Re_p}) \left(1 + \alpha Re_p^{\beta}\right)
\end{equation}
with $\alpha=0.1315$, $\beta=0.82-0.05\,log_{10}Re_p$ when $Re_p < 20$, and $\alpha=0.1935$, $\beta=0.6305$ for $Re_p \ge 20$.
Using the definitions of $F_D^S$ and $C_{D_0}$ in equation~(\ref{eq_sph_av}) we obtain
\begin{equation}
\label{eq_sph4}
\frac{4}{3} \pi a^3 (\rho_p - \rho_f) g = \frac{1}{2} \rho_f \pi a^2 \langle |\vec U_{r}| U_{r,z} \frac{24}{Re_p}\left(1 + \alpha Re_p^{\beta}\right) \rangle+ F_D^U.
\end{equation}
Next, we define the new variable $K=\alpha Re_p^{\beta}$ and divide equation~(\ref{eq_sph4}) by the buoyancy term to find
\begin{equation}
\label{eq_sph6}
1 = \frac{\langle U_{r,z} (1+K)\rangle}{V_s} + f_D^U,
\end{equation}
where $V_s=2(\rho_p - \rho_f)g a^2/(9 \rho_f \nu)$ is the Stokes settling velocity and $f_D^U=F_D^U/(4\pi a^3(\rho_p-\rho_f)g/3)$. 

The steady nonlinear term, the first on the right-hand side above, can be split into two components by 
writing  the relative velocity along gravity and $K$ as the sum of 
their mean values and fluctuations: $U_{r,z} = \overline U_{r,z} + U_{r,z}' $ and $K = \overline K + K'$ (where 
$\overline K = \langle \alpha Re_p^{\beta}\rangle$ and $K'$ are the fluctuations about the mean value $\overline K$). Equation~(\ref{eq_sph6}) hence 
becomes
\begin{equation}
\label{eq_sph7}
1 = \frac{\overline U_{r,z} (1 + \overline K)}{V_s} + \frac{\langle {U_{r,z}'\,K'}\rangle}{V_s}  + f_D^U
\end{equation}
The term corresponding to the contribution from mean settling in equation~(\ref{eq_sph7}) can be further decomposed by defining $\widehat K = \alpha 
\left(2a\overline U_{r,z}/\nu \right)^{\beta}$, such that $\overline K = \widehat K + K''$. Note that $\overline K$ is calculated 
with the particle Reynolds number defined with $|\vec U_r|$, and if the particle lateral motions vanish, $\widehat K$ coincides with $\overline K$ 
(i.e. $K'' \approx 0$). In contrast, when particles are horizontally swept by turbulent eddies the term $K''$ becomes greater than zero. 
Thus we finally get
\begin{equation}
\label{eq_sph8}
1 = \frac{\overline U_{r,z} (1 + \widehat K)}{V_s} + \frac{\overline U_{r,z}\,K''}{V_s} + \frac{\langle U_{r,z}'\,K'\rangle}{V_s} + f_D^U,
\end{equation}
where $\overline U_{r,z} (1 + \widehat K)$ is the \emph{drag of the mean settling} and is the only term in the case of a single sphere settling in still fluid.
$\overline U_{r,z}\,K''$ 
is the \emph{cross-flow-induced drag}, i.e.\ the vertical component of the drag due to a non-zero horizontal relative motion. We name the term $\langle {U_{r,z}'\,K'}\rangle$ as 
\emph{nonlinear-induced drag.} This term becomes relevant in a turbulent flow  when the variance of the particle velocity is larger than in still fluid. Finally, 
the term $f_D^U$ accounts for unsteady effects, particle-particle and wake interactions and increases with the volume fraction \citep{fornari2015}.

\begin{figure}
  \centering
  \subfigure{%
    \includegraphics[scale=0.35]{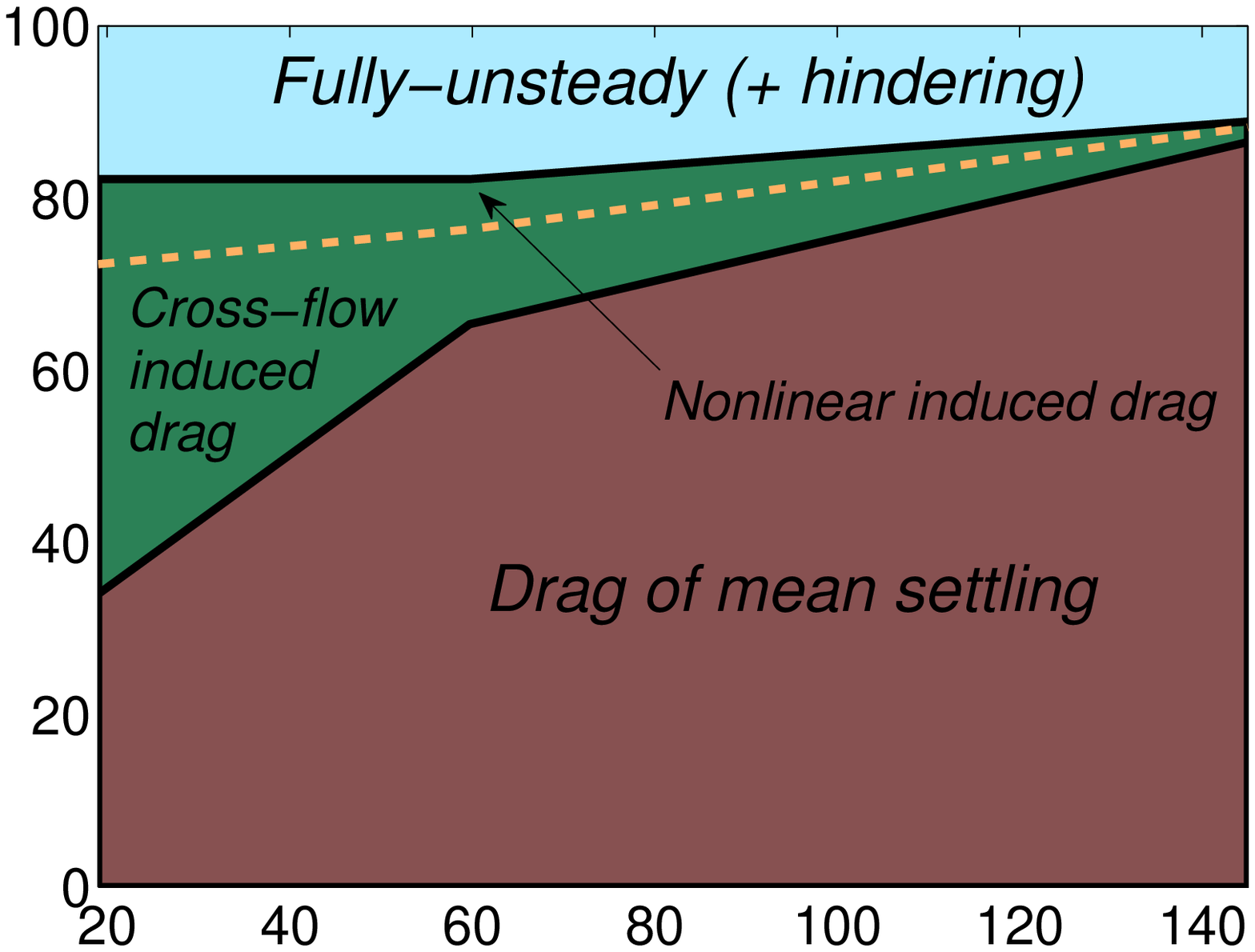}
     \put(-194,130){{\large a)}}
     \put(-194,65){\rotatebox{90}{\large $\%$}}
     \put(-94,-5){{\large $Ga$}}
     }%
  \subfigure{%
    \includegraphics[scale=0.35]{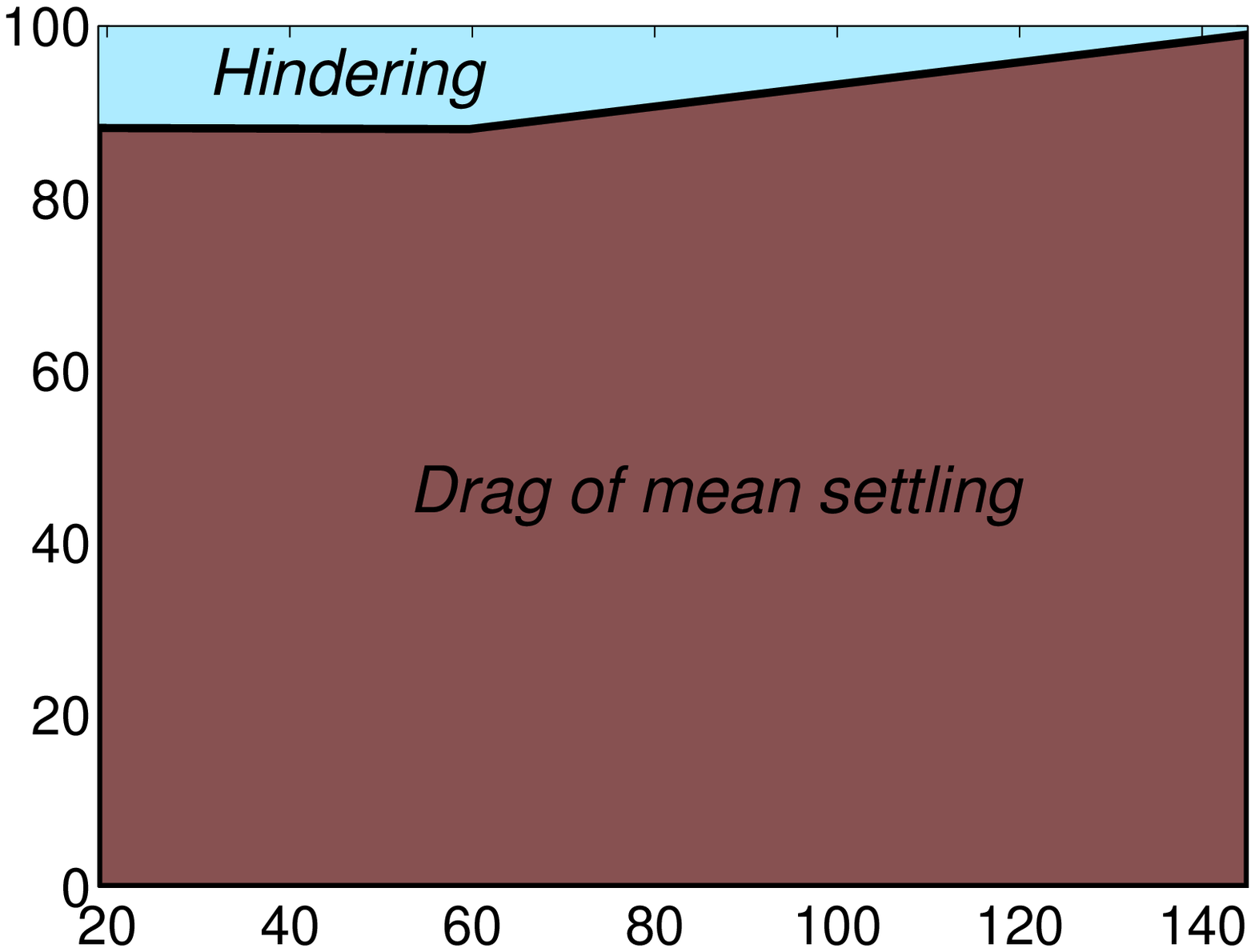}   
     \put(-194,130){{\large b)}}
     \put(-194,65){\rotatebox{90}{\large $\%$}}
     \put(-94,-5){{\large $Ga$}}
}%
\caption{Relative contributions to the overall mean drag: (a) turbulent flow; (b) quiescent environment. The dashed line separates the contributions from the 
\emph{nonlinear induced drag} and the \emph{cross-flow induced drag}.}
\label{fig:dragmap}
\end{figure}

The relative importance of the four terms on the right-hand side of equation (\ref{eq_sph8}) is shown as percentage of the total drag in figure~\ref{fig:dragmap}(a,b) 
for settling in turbulent and quiescent flow. 
In turbulence, the contributions 
due to the nonlinear-induced drag and the cross-flow-induced drag increase while reducing $Ga$. 
These increase from $1\%$ and $2\%$ 
at $Ga=145$ to $9\%$ and $38\%$ at $Ga=19$ and are the main responsible for the reduction in settling speed shown above. 
For $Ga=19$ the mean nonlinear drag is $44\%$ of the overall steady drag when settling occurs in quiescent fluid, 
while it increases to $72\%$ of the overall mean steady drag in a turbulent flow. 
Thus, at small $Ga$, the nonlinear part of the overall mean steady drag increases drastically in turbulence. 
When $Ga$ is large, in contrast, the nonlinear contribution to the overall mean steady drag is of the same order in the quiescent and turbulent cases (e.g. for $Ga=145$ it 
amounts to $84\%$ of the mean steady drag in both cases).

Unsteady effects also increase the overall drag in turbulent flows, see 
figure~\ref{fig:dragmap}(a). 
The increase of the unsteady contribution in turbulent flow amounts to about $10\%$ for $Ga=145$ and $6\%$ for $Ga=19$ and $60$, 
which alone is not enough to explain the 
large reduction in $\langle V_z \rangle_{p,t}$ at the lower $Ga$. These values are an estimate obtained by subtracting the 
percentage obtained in the quiescent cases to the percentage for the turbulent cases and show the additional increase in drag due to new unsteady 
effects, clearly weaker for particles settling in a quiescent fluid.

Further, we investigate the contribution of added mass to the overall mean drag. The added mass force is expressed as 
$\chi \left(4\pi \rho_f a^3/3 \right) \left(\md{U_z} - \td{V_z}{t} \right)$, where $\md{U_z}$ is the material derivative of the fluid velocity seen by the particle, 
and $\chi=0.5$ the added mass coefficient~\citep{chang1995,merle2005}. For each particle, we calculate the added mass force as an average over 
the volume of the spherical shells used to estimate the relative velocities. Finally we compute an ensemble average for each particle and all time steps and normalize by the buoyancy term. 
The average values indicate that for the 
larger Galileo numbers the added mass contribution is negligible. For the lower $Ga$ values, we find instead that the added mass contributes by about $1.5\%$ of the 
overall mean drag and $25\%$ of the unsteady term.

Hence, our main finding is 
that when the turbulent velocity fluctuations $u'$ are larger than the characteristic reference 
velocity of the settling process (i.e. the terminal velocity of an isolated particle $V_t$)
the overall drag significantly increases. This is due to the increased intensity of the particle relative motions 
and to the increase of the variance of the particle velocity. 

\begin{figure}
  \centering
  \subfigure{%
    \includegraphics[scale=0.35]{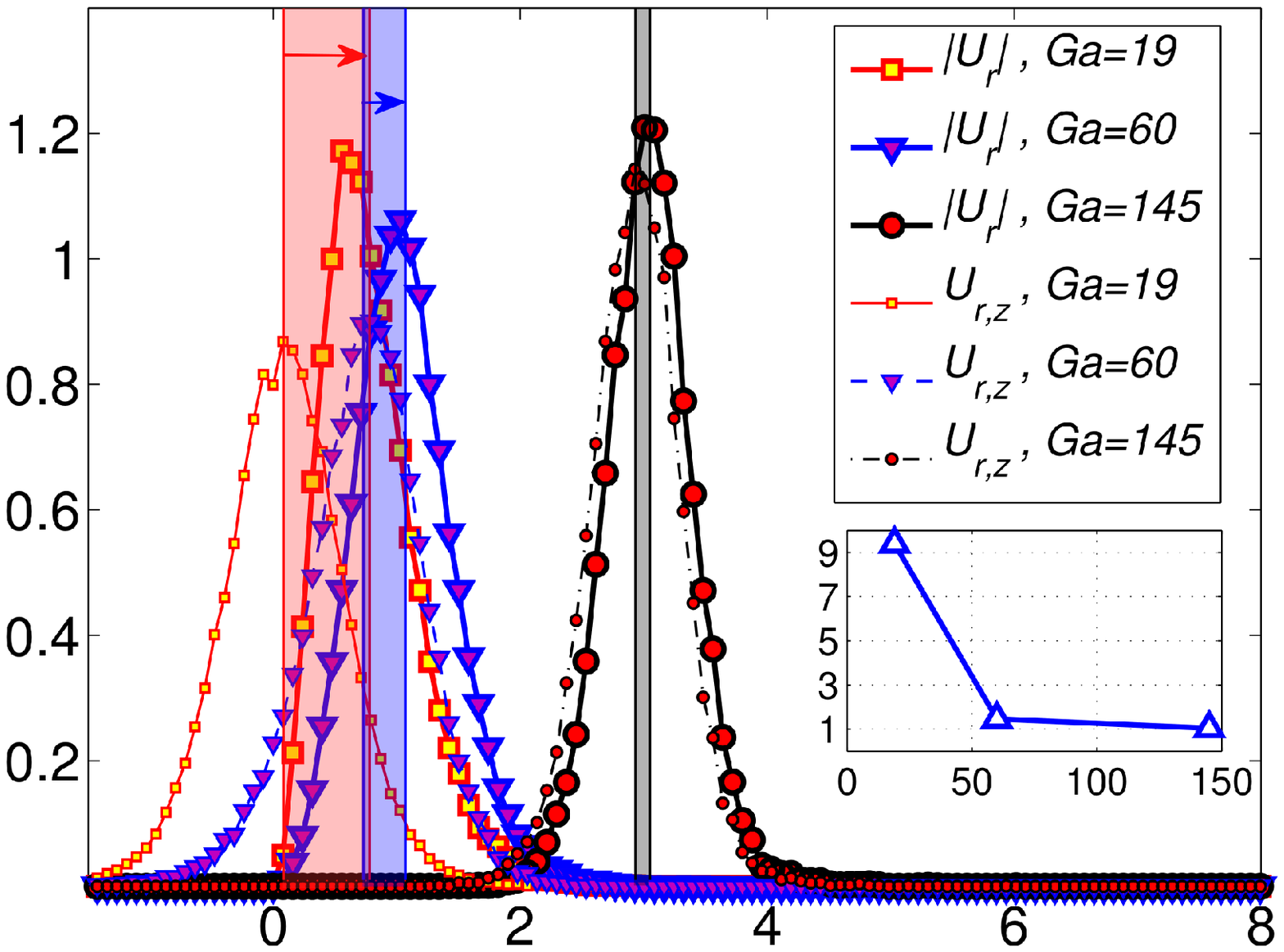}
     \put(-194,130){{\large a)}}
     \put(-196,60){\rotatebox{90}{\large $p.d.f.$}}
     \put(-132,-5){{\large $|\vec U_r|/u'$ , \, $U_r/u'$}}
     \put(-55,50){\scriptsize $\overline{|\vec U_r|}/\overline U_r$}
     \put(-50,20){\scriptsize $Ga$}
     }%
  \subfigure{%
    \includegraphics[scale=0.35]{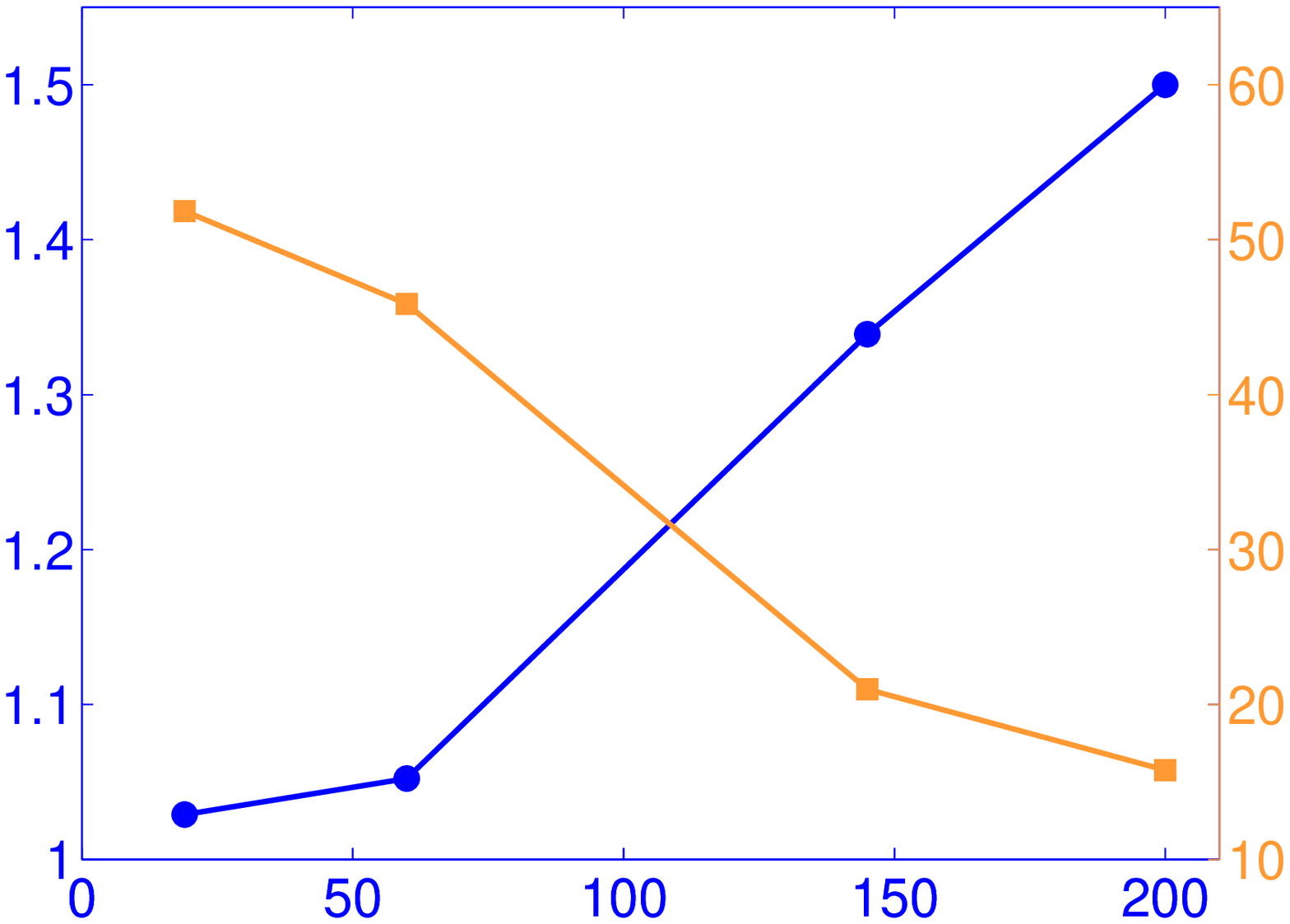}   
     \put(-194,130){{\large b)}}
     \put(-198,65){\rotatebox{90}{\large $\frac{\sigma_n}{\sigma_z}$}}
     \put(-32,65){\rotatebox{90}{\large $\sigma_{\theta^\circ}$}}
     \put(-96,-5){{\large $Ga$}}
}%
\caption{(a) Probability density functions, $p.d.f.$s, of $|\vec U_r|$ and $U_{r,z}$. The shaded zones show the 
difference between the mean values of $|\vec U_r|$ and $U_{r,z}$. The inset shows the ratio between 
$\overline {|\vec U_r|}$ and $\overline U_{r,z}$. (b) Anisotropy in the particle velocity fluctuations 
$\sigma_n/\sigma_x$ (blue circles) and standard deviation of the settling angle $\sigma_{\theta^\circ}$ (orange squares) 
as a function of $Ga$.}
\label{fig:dragmap1}
\end{figure}
In quiescent fluid, the major contribution to the overall drag comes from the term $\overline 
U_{r,z} (1 + \widehat K)$, see figure~\ref{fig:dragmap}(b). The remaining part is due to the hindering effect ($1\%$ for $Ga=145$ and 
approximately $10\%$ at lower $Ga$). These values are in agreement with the reduction of the mean settling velocity $\langle V_z \rangle_{p,t}$ reported above.
Note that we checked that for an isolated particle 
in quiescent fluid $F_D^S = 4 \pi a^3 (\rho_p - \rho_f) \vec g/3$ (within $\pm 3.5\%$ error for all $Ga$). Indeed, for single particles at these $Ga$ there are no 
unsteady contributions to the overall drag and $F_D^U=0$. Possible reasons for this inaccuracy are the use of empirical formulae, and the thickness of 
the shell used in the definition of $\vec U_r$.

Given the finding above, we quantify the importance of the cross-flow induced motions 
by examining the $p.d.f.$s of $|\vec U_r|$ and $U_{r,z}$, shown in figure~\ref{fig:dragmap1}(a).
First, we consider the origin of possible differences between the absolute relative velocity and the mean settling speed. 
To this aim, we express each velocity component in $|\vec U_{r}|$ as the sum of mean and fluctuations,
\begin{equation}
|\vec U_{r}| = \sqrt{U_{r,x}'^2+U_{r,y}'^2+(\overline U_{r,z} + U_{r,z}')^2}= \overline U_{r,z} \sqrt{\frac{U_{r,x}'^2}{\overline U_{r,z}^2} + \frac{U_{r,y}'^2}{\overline U_{r,z}^2} + \left(1+\frac{2U_{r,z}'}{\overline U_{r,z}}+\frac{U_{r,z}'^2}{\overline U_{r,z}^2}\right)}
\end{equation}
(with $\overline U_{r,x} \simeq \overline U_{r,y} \simeq 0$ by symmetry). In the quiescent cases $U_{r,i}' \ll \overline U_{r,z}$ and 
$|\vec U_{r}| \simeq \overline U_{r,z}$. 
In a turbulent flow, 
if $V_t/u'$ is large the quadratic terms are negligible and 
$|\vec U_{r}| \simeq \overline U_{r,z} \sqrt{1+2U_{r,z}'/\overline U_{r,z}}$ (i.e. only fluctuations in the direction of gravity are 
important). If $V_t/u'<1$ , instead, the quadratic terms are dominant and $|\vec U_{r}|$ can be substantially larger than $\overline U_{r,z}$.

We see indeed in figure~\ref{fig:dragmap1}(a) that the difference $\overline{|\vec U_{r}|}-\overline U_{r,z}$ (given by the shaded zones) 
increases by reducing $Ga$. 
Roughly assuming that $U_{r,i}' / \overline U_{r,z}\approx u'/V_t$, one can obtain estimates of the relative velocity $|\vec U_{r}|$ in agreement with the DNS data:
for $Ga=19$, we obtain $\overline{|\vec U_{r}|} \sim 9.3 \overline U_{r,z}$ instead of the actual value $9.4$, 
for $Ga=60$, $\overline{|\vec U_{r}|} \sim 2.4 \overline U_{r,z}$ instead of $1.5$, 
for $Ga=145$, $\overline{|\vec U_{r}|} \sim 1.3 \overline U_{r,z}$ instead of $1$, 
see the inset of figure~\ref{fig:dragmap1}(a). Hence, the cross-flow-induced drag may be estimated a priori.
It is important to note that at low $Ga$, the relative velocity fluctuations in the direction of gravity also 
contribute to $|\vec U_{r}|$ and therefore to the increase in drag. However, the contribution from 
transverse fluctuations is twice that in the direction of gravity. Hence the name \emph{cross-flow-induced drag}.

Here $\langle |\vec U_{r}| \rangle$ can be used to estimate  the particle Reynolds number $Re_p=(2a)\langle |\vec U_{r}| \rangle/\nu$ for each case studied. 
For $Ga=19$, $60$, $145$ and $200$ we find $Re_p = 45.6$, $61.9$, $176.7$ and $263.6$ in turbulence, and $Re_p = 9.3$, $52.1$, $185.5$ and $272.4$
in quiescent fluid. This confirms that the relative velocities drastically increase at low $Ga$ in the turbulent cases.

In figure~\ref{fig:dragmap1}(b) we show the ratio between the particle fluctuating velocities in the directions perpendicular and parallel to 
gravity, $\sigma_n/\sigma_z$. We note that $\sigma_n/\sigma_z$ grows with $Ga$. At the lowest $Ga$, the particle velocity fluctuations are 
approximately the same in all directions. At high $Ga$ the anisotropy increases ($\sigma_n/\sigma_z$ increases), as particles fall faster and 
the fluctuations in the vertical direction are relatively less important. As $\sigma_n/u'$ is approximately constant for all cases, particles 
are seen to undergo less intense lateral motions at high $Ga$ when the dynamics is  dominated by buoyancy. Interestingly, the 
opposite behaviour is  observed for heavy point-particles \citep{good2014} showing the importance of particle size.
%
In the same figure, we also display the standard deviation $\sigma_{\theta^\circ}$ of the angle between the mean particle velocity and 
the vertical axis, $\theta = arctg(V_n/V_z)$. In agreement with the previous observations, $\sigma_{\theta^\circ}$ is $4$ times 
larger for $Ga=19$ than for $Ga=200$. Heavy finite-size particles fall along almost straight vertical paths, whereas lighter particles are 
strongly swept laterally by intense eddies and, hence, the larger $\sigma_{\theta^\circ}$.

\begin{figure}
  \centering
  \subfigure{%
    \includegraphics[scale=0.35]{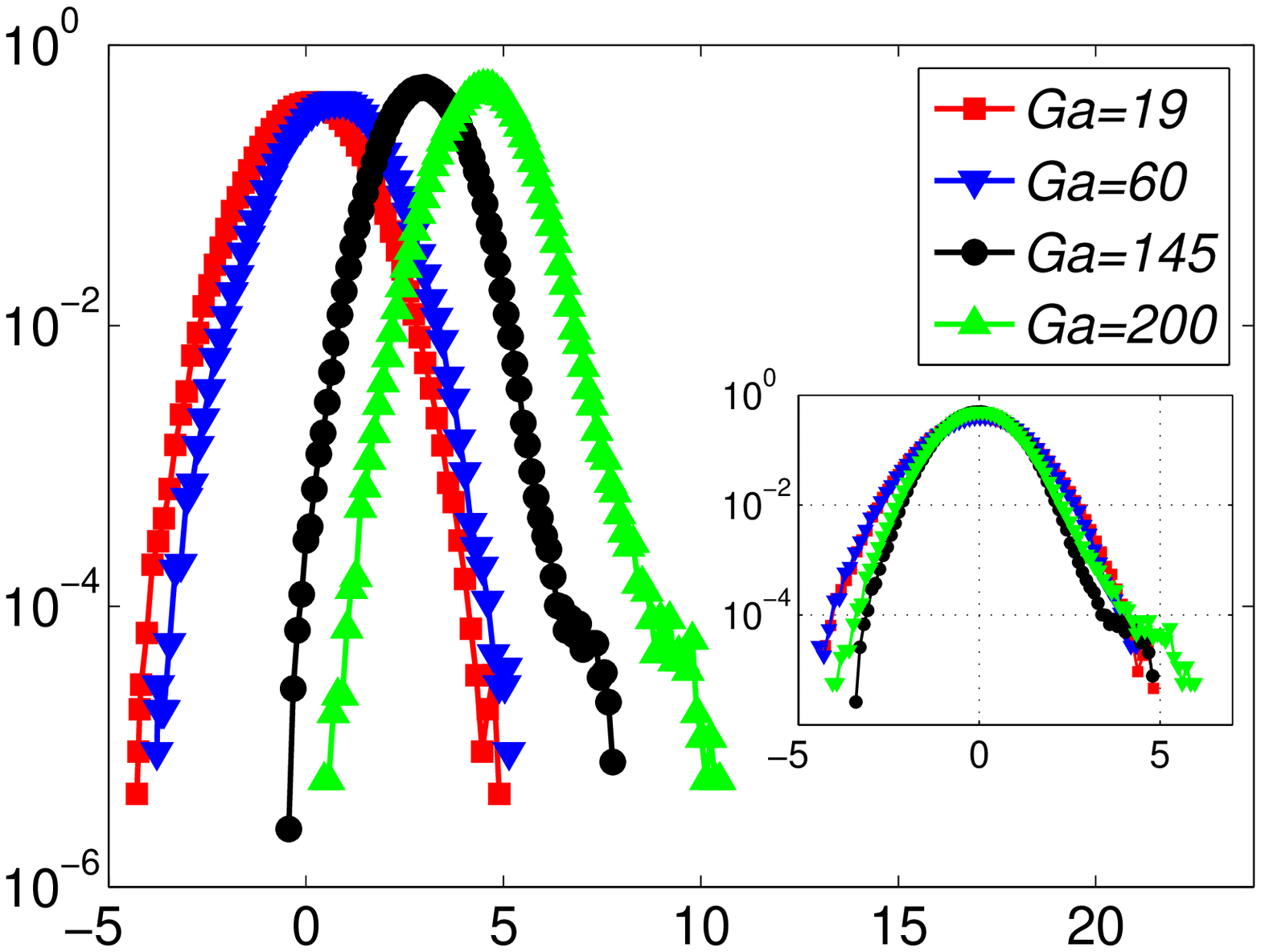}
     \put(-194,130){{\large a)}}
     \put(-196,63){\rotatebox{90}{\large $p.d.f.$}}
     \put(-106,-5){{\large $V_z/u'$}}
     \put(-98,65){\rotatebox{90}{\scriptsize $p.d.f.$}}
     \put(-70,25){\scriptsize $V_z - \langle V_z \rangle_{p,t}$}
     }%
  \subfigure{%
    \includegraphics[scale=0.35]{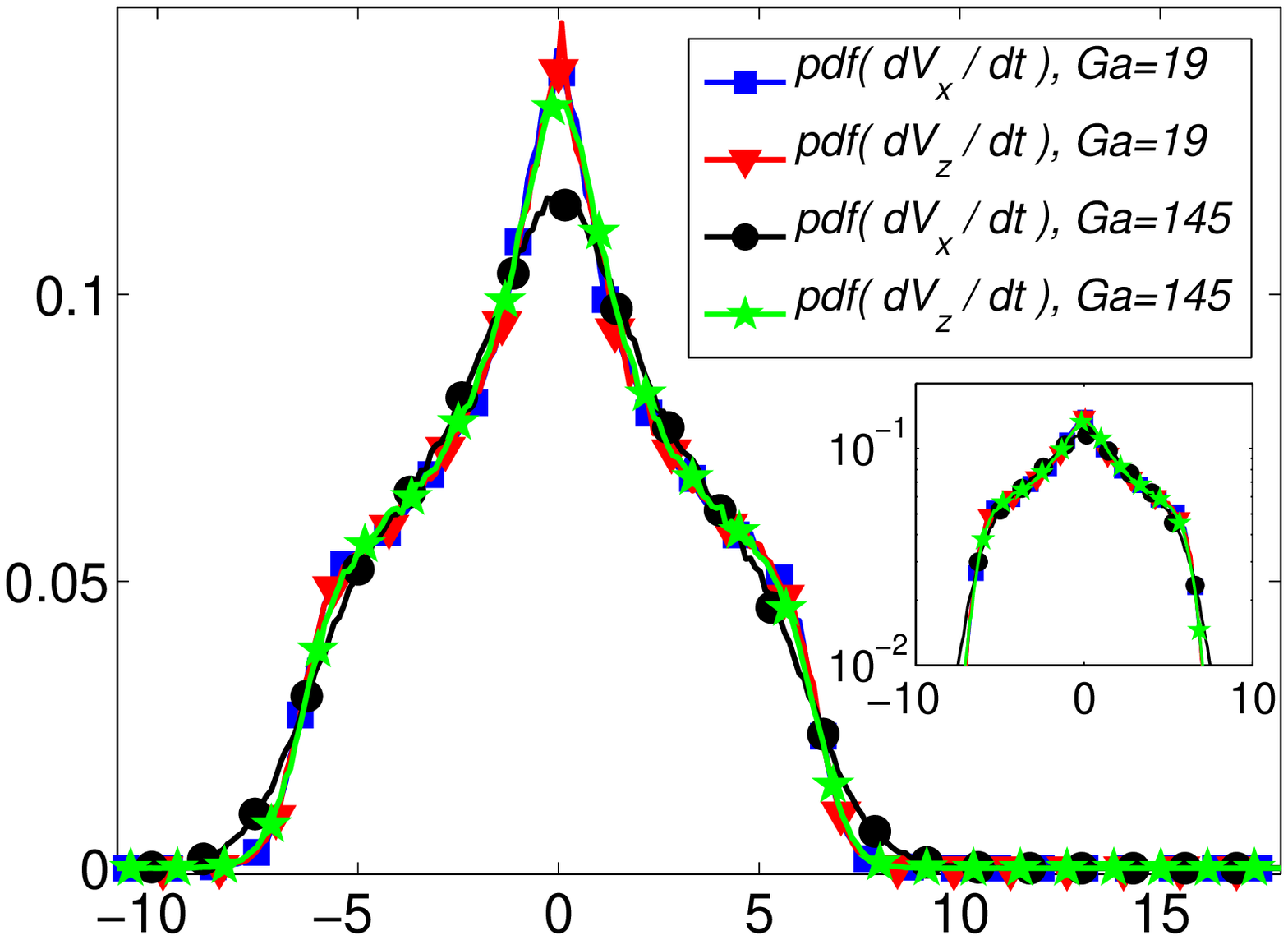}
     \put(-194,130){{\large b)}}
     \put(-196,63){\rotatebox{90}{\large $p.d.f.$}}
     \put(-134,-5){{\large [$dV_i/dt] \, [(2a)/u'^2]$}}
     }%
\\
  \subfigure{%
    \includegraphics[scale=0.35]{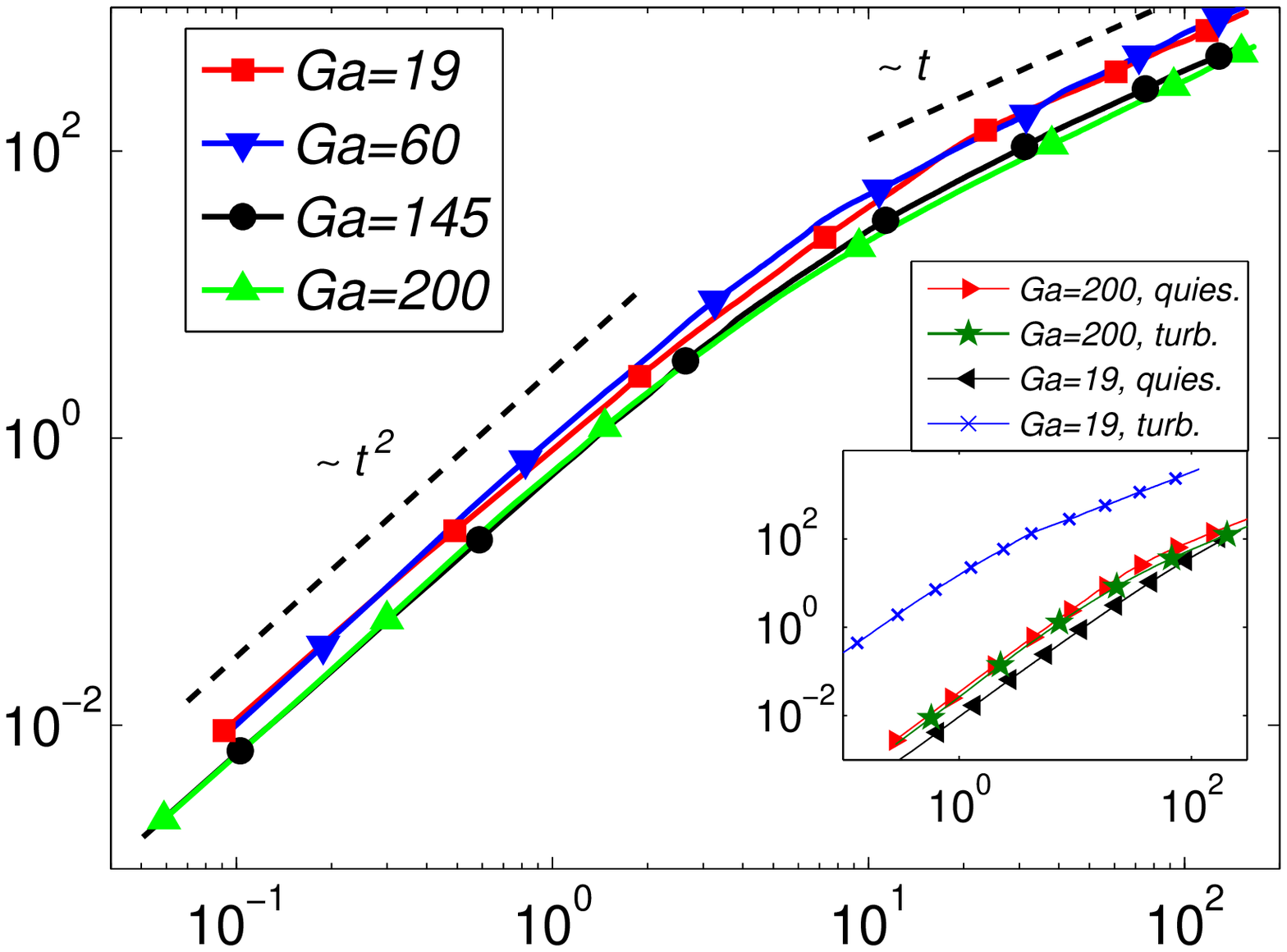}   
     \put(-194,130){{\large c)}}
     \put(-201,65){\rotatebox{90}{\large $S_2^Z$}}
     \put(-98,44){\rotatebox{90}{\scriptsize $S_2^Z$}}
     \put(-53,34){{\scriptsize $t V_t/(2a)$}}
     \put(-116,-5){{\large $t u'/(2a)$}}
}%
\caption{(a) Probability density functions, $p.d.f.$s, of $V_z/u'$ and $\left(V_z - \langle V_z \rangle_{p,t}\right)/u'$ (inset) in turbulence. (b) 
Probability density functions, $p.d.f.$s, of particle accelerations in turbulence ($dV_x/dt$ and $dV_z/dt$, scaled by $(2a)/u'^2$), for $Ga=19$ and $145$. 
(c) Particle mean-square displacement in the settling direction $S_2^Z$ in turbulence. In the inset we show the mean-square displacements along the 
settling direction for $Ga=19$ and $200$, for both quiescent and turbulent cases.}
\label{fig:pdfvel}
\end{figure}

As discussed above, the particle mean settling speed $\langle V_z \rangle$ is mostly governed by buoyancy at high $Ga$, while it is affected strongly by turbulence 
at low $Ga$. It is, however, interesting to observe that the variance of the particle settling speeds is similar for each $Ga$. To show this we report in 
figure~\ref{fig:pdfvel}(a) the $p.d.f.$s of particle settling speeds $V_z$ and of $V_z - \langle V_z \rangle_{p,t}$ (inset) in turbulence, 
normalized by $u'$. these indicate that the variance is similar in all cases and that the different curves almost overlap. Hence, 
although the effect of the turbulence on the mean settling speed depends on $V_t/u'$, fluctuations of the settling speed depend mostly on the properties 
of the turbulent flow (i.e. on $u'$).

Next we examine particle accelerations in turbulence. We therefore depict in figure~\ref{fig:pdfvel}(b) the $pdf$s of $dV_x/dt$ and $dV_z/dt$ 
scaled by $u'^2/(2a)$, for $Ga=19$ and $145$ (since the configuration is axialsymmetric $dV_y/dt$ is not reported). We note, surprisingly, that the $pdf$s 
of particle accelerations collapse onto one curve (the $pdf$s are not shown for $Ga=60$ and $200$ for the sake of clarity); only the $p.d.f.$s of 
$dV_x/dt$ for $Ga=145$ and $200$ are slightly different. Hence particle accelerations are also almost completely governed by turbulence and all curves 
are almost perfectly symmetric. Note also that similar shapes for the $pdf$s of accelerations have been found experimentally for negatively buoyant spheres \citep{mathai2015}.

The particle mean-square displacement in the settling direction $S_2^Z=\langle \left(\Delta z - \langle V_z \rangle t \right)^2 \rangle$ is also mainly determined
by the properties of the turbulent flow. As shown in figure~\ref{fig:pdfvel}(c), 
the  mean-square displacements pertaining to all $Ga$ almost collapse when scaling time with $(2a)/u'$. In the inset of the same figure, we compare $S_2^Z$ for 
quiescent and turbulent cases at $Ga=19$ and $200$, where times are scaled by $(2a)/V_t$. For particles falling relatively fast, $Ga=200$, 
$S_2^Z$ is similar in quiescent and turbulent flow. In contrast,  $S_2^Z$ increases significantly in the turbulent case for the smallest $Ga$ under investigation.
 
To conclude, we would like to note that our results are qualitatively consistent with the findings of \citet{homann2013} 
and \citet{chouippe2015}. These authors suggest that the nonlinear drag acting on particles in a turbulent flow is higher than in laminar or 
quiescent fluid. They suggest that the relative increase in drag is proportional to $C(Re_p)I^2$, where $C(Re_p)$ is a nonlinear function of the 
particle Reynolds number $Re_p$, while $I \sim u'/U_{r,z}$ is the relative turbulence intensity (similar to our definition). This scaling confirms 
that drag nonlinearity increases as the relative turbulence intensity $I$ increases, in agreement with our results.

\subsection{Comparison with sub-Kolmogorov particles}

\begin{figure}
  \centering
  \subfigure{%
    \includegraphics[scale=0.3]{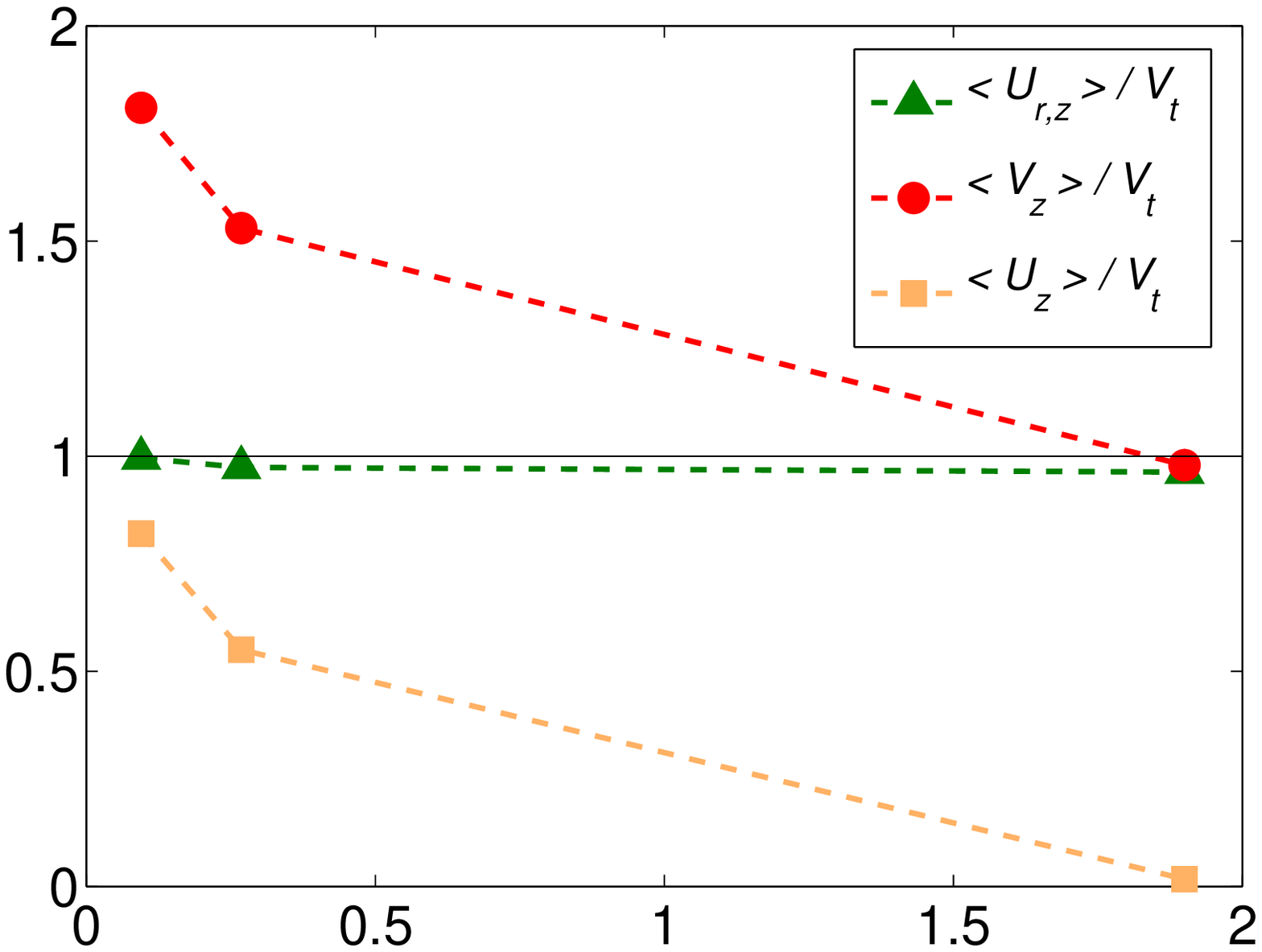}
     \put(-170,100){{\large a)}}
     \put(-170,65){\rotatebox{90}{\large $V$}}
     \put(-94,-5){{\large $V_t/u'$}}
     }%
  \subfigure{%
    \includegraphics[scale=0.3]{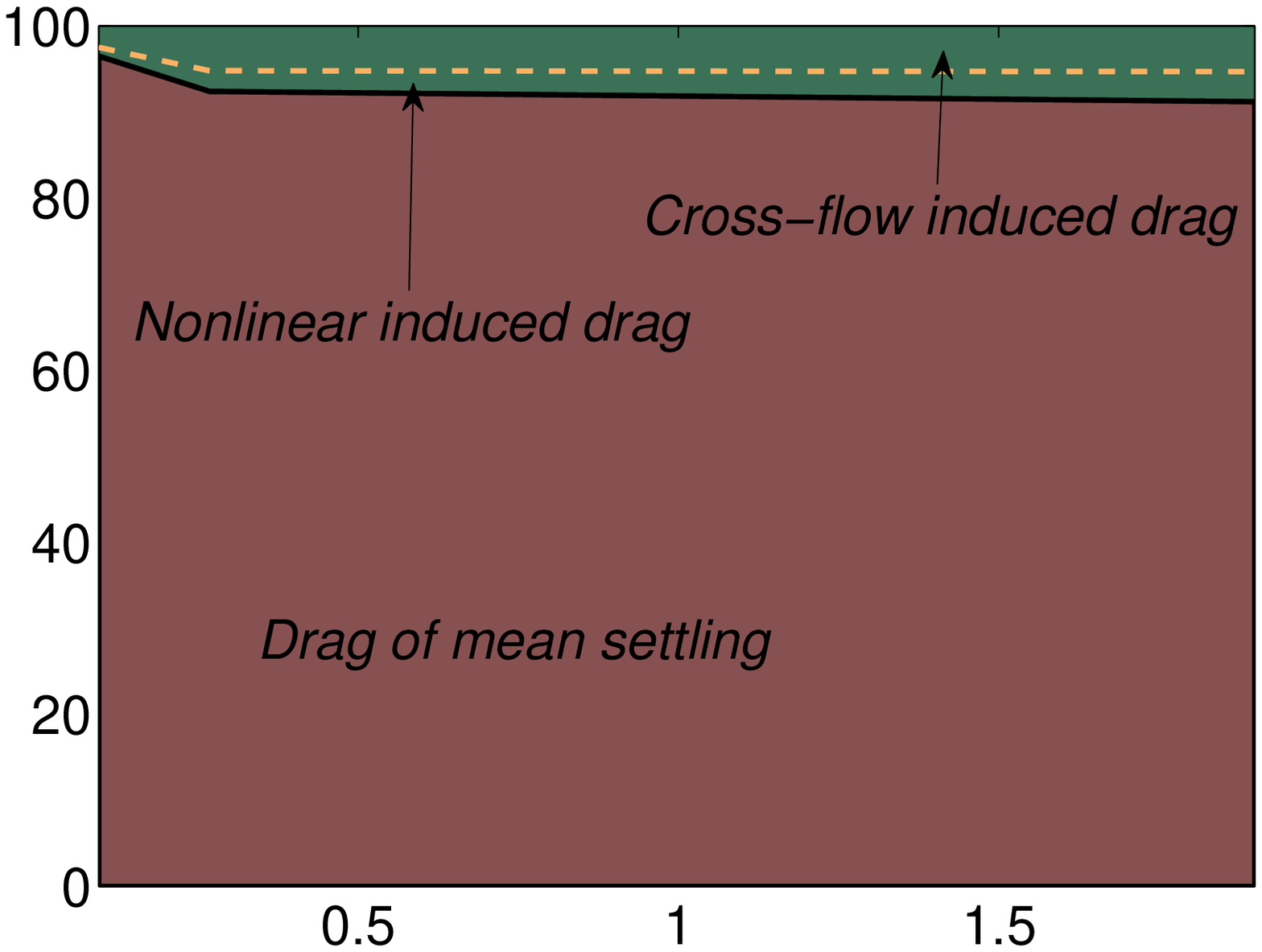}   
     \put(-170,100){{\large b)}}
     \put(-170,65){\rotatebox{90}{\large $\%$}}
     \put(-94,-5){{\large $V_t/u'$}}
}%
\caption{(a) $\langle V_{z}\rangle/V_t$, $\langle U_{z}\rangle/V_t$ and $\langle U_{r,z}\rangle/V_t$ and  
(b) Drag map for the point-particle cases as function of $V_t/u'$. $V_s/u'=0.1$, $0.3$, $3$ correspond to $V_t/u'=0.095$, $0.268$, $1.9$.}
\label{fig:dragpoint1}
\end{figure}

Finally, we wish compare the reduction observed for finite-size slightly buoyant particles to the behaviour of heavy sub-Kolmogorov particles. 
Their settling speed, $\langle V_z \rangle/V_t$, is found to reduce when increasing particle inertia (i.e. increasing $V_t/u'$ and $\tau_p$), the opposite of what reported here for finite-size particles.
We therefore performed simulations of heavy ($\rho_p/\rho_f \sim 1000$) point-particles in HIT  \citep[see][for details on the method]{olivieri}. 
At high $\rho_p/\rho_f$, the particle acceleration is determined only by gravity and drag 
\begin{equation}
\label{eq_p0}
\td{\vec V}{t} = -\frac{\vec U_{r}}{\tau_p} \zeta + \vec g 
\end{equation}
where $\zeta = 1 + 0.15Re_p^{0.687}$ is the nonlinear drag correction also used by \citet{good2014}. 
We study three different cases, characterized by  particle relaxation times $\tau_p = 0.389, 1.296$ and $12.96$ and ratio between settling speed and turbulence intensity $V_s/u' = 0.1, 0.3$ and $3$. The turbulent flow field has 
$Re_{\lambda}=90$ and a ratio $u'/u_{\eta}=4.77$.

in figure~\ref{fig:dragpoint1}a)  we report the mean settling speed, $\langle V_z \rangle/V_t$ where the reference settling speed is $V_t=V_s/\zeta$, 
the mean fluid velocity sampled by the  particles $\langle U_z \rangle$ and the relative velocity $\langle U_{r,z}\rangle/V_t$. 
In agreement with the previous studies mentioned above, the mean settling speed decreases with $V_t/u'$ and becomes smaller than 
unity for the largest $V_t/u'$ ($\tau_p=12.96$). The fluid velocity at the particle position is positive at small $V_t/u'$ (small $\tau_p$) and tends 
to zero when increasing the particle inertia. 
In other words, at small $V_t/u'$, preferential sweeping occurs and sub-Kolmogorov particles settle with $\langle V_z \rangle > V_t$.
The reduction of the mean settling velocity at large $V_t/u'$ is thus due to the absence of sampling of downdrafts. 

It can be proven that using nonlinear drag corrections is necessary to find $\langle V_z \rangle/V_t < 1$, as observed at large $V_t/u'$. 
To this end, we express $U_{r,z}$ and $V_z$ in terms of a power series of the modified Reynolds number $\widehat Re=0.15Re_p^{0.687}$: 
\begin{equation}
U_{r,z} = u_0 + u_1 \widehat Re + u_2 \widehat{Re}^2 + h.o.t 
\end{equation}
\begin{equation}
V_{z} = v_0 + v_1 \widehat Re + v_2 \widehat{Re}^2 + h.o.t. 
\end{equation}
where $h.o.t.$ denotes higher order terms. Substituting in the particle equation~(\ref{eq_p0}), projecting along gravity and time averaging, we obtain 
at first order  
\begin{equation}
\langle V_z \rangle = \langle U_z \rangle + V_t \left( 1 - \widehat{Re}^2 \right).
\end{equation}
It appears therefore clearly that drag nonlinearity is responsible for the reduction of settling speed in the absence of preferential sampling. 

The effect of drag nonlinearity is however modest and the nonlinear contribution to the overall drag 
increases slowly with  $V_t/u'$. To show this, we perform the same force analysis done for finite-size particles. 
We average equation~(\ref{eq_p0}) and express velocity and drag coefficient as the sum of mean values ($\overline{\cdot}$) and fluctuating components ($\cdot'$). 
As above, we finally obtain
\begin{equation}
\label{eq_p3}
g \tau_p = \overline U_{r,z} (1 + \widehat K) + \overline U_{r,z} K'' + \langle {U_{r,z}'\,K'}\rangle
\end{equation}
with $\overline K = \widehat K + K''$ (where $\widehat K = 0.15\left(2a\overline U_{r,z}/\nu \right)^{0.687}$). 
The results in figure~\ref{fig:dragpoint1}b) show that increasing $V_t/u'$, the nonlinear-induced 
drag and the cross-flow-induced drag only slightly increase.

We finally consider the role of loitering \citep{nielsen1993}. As in 
\citet{good2014} we perform simulations in which the particles are constrained to settle along vertical paths. In such artificial case, 
the settling speed, $\langle V_{p,z}\rangle/V_t$, is always less than one and the fluid velocity at the particle position, 
$\langle U_{z}\rangle/V_t$, is negative (opposite to the direction to gravity). In other words, particles preferentially sample updrafts. 
According to \citet{nielsen1993}, loitering 
should become effective when $V_s/u' \ge 0.3$. However, simulations of particles free to move horizontally show that the preferential sampling occurs always in downdrafts and tends to zero as $V_t/u'$ (or equivalently $\tau_p$) increases.
Loitering seems therefore absent in HIT. 

\section{Final remarks}

We have examined how the settling of finite-size particles in HIT is affected by the relative turbulence intensity, 
$V_t/u'$. In particular, we have considered particles that are slightly heavier than the carrier fluid.
We find that as $Ga$ or $V_t/u'$ decrease the mean settling speed reduces. This reduction is stronger when $V_t/u' < 1$ (about $60\%$ for $Ga=19$ 
and $V_t/u'=0.19$). We attribute this reduction at small $V_t/u'$ to unsteady effects and, prominently, to drag nonlinearity. 
The component of the drag acting in the direction of gravity increases due to the increase of the fluid-particle relative velocity in comparison 
to the quiescent cases. This is associated with the strong lateral/cross-flow motions induced by the turbulent eddies on the settling particles. 
When $Ga$ (or $V_t/u'$) is large, particles fall along vertical paths with weak relative lateral velocities. Lighter particles, instead, are subjected 
to velocity fluctuations one order of magnitude larger than their mean settling velocities. These particles have therefore significant lateral relative 
velocities and this, in turn, determines an increase of the drag acting on them. Indeed at small $Ga$, the particle Reynolds number $Re_p$ 
(defined via the magnitude of the slip 
velocity $|\vec U_r|$) is substantially larger in a turbulent flow than in quiescent fluid, for which it is of the order of the average settling speed $\langle |\vec U_r| \rangle \sim \langle V_z \rangle$.

Although the reduction of the mean settling speed, $\langle V_z \rangle$, depends on the relative turbulence intensity (i.e. $V_t/u'$), other 
quantities depend exclusively on the turbulent velocity fluctuations, $u'$. Indeed, we report that the particle velocity fluctuations 
are very similar in all directions  for all $Ga$. These are just slightly smaller in the direction perpendicular to gravity at the higher $Ga$. Moreover,  the $pdf$s of particle accelerations and the mean-square displacements almost perfectly collapse on a single curve 
when scaling appropriately the different quantities with the turbulence intensity, $u'$.

Finally, we have compared the behaviour of finite-size and  sub-Kolmogorov particles. As already discussed, the mean settling speed of large particles 
decreases as their inertia is reduced (i.e. lower $Ga$ and $\tau_p$). The reduced settling speed of  finite-size particles is related to the 
large fluctuations of the relative velocity and, hence, to the increase of the nonlinear drag. 
In contrast, the variation of settling speed of point-like particles can be mainly explained by preferential sampling of downdrafts. 
Indeed, for this type of particles the mean settling speed is reduced as their inertia increases and eventually reaches a plateau around $\langle V_z 
\rangle \simeq V_t$ at very large $\tau_p$, since preferential sampling progressively disappears. For the range of parameters here studied, we 
also find that loitering plays a negligible role in reducing the mean settling speed of point-like particles.

%
\begin{acknowledgments}
This work was supported by the European Research Council grant no.\ ERC-2013-CoG-616186, TRITOS and by the Swedish Research Council (VR).
Computer time was provided by SNIC (Swedish
National Infrastructure for Computing). The support from the COST Action MP1305: \emph{Flowing matter} is also acknoledged.
\end{acknowledgments}

\bibliographystyle{jfm}

\bibliography{jfm-instructions}

\end{document}